\documentclass[a4paper,fleqn]{cas-dc}

\usepackage[numbers,compress]{natbib}
\usepackage{graphicx}
\usepackage[T1]{fontenc}

 \usepackage{microtype}
\def\tsc#1{\csdef{#1}{\textsc{\lowercase{#1}}\xspace}}
\tsc{WGM}
\tsc{QE}
\tsc{EP}
\tsc{PMS}
\tsc{BEC}
\tsc{DE}

\def\vec#1{\boldsymbol{#1}}

\def\llcol#1#2{\tilde{\lambda}_{#1}.\tilde{\lambda}_{#2}}
\def\llss#1#2{\tilde{\lambda}_{#1}.\tilde{\lambda}_{#2}\,\boldsymbol{\sigma}_{#1}.\boldsymbol{\sigma}_{#2}}

\begin{document}
\let\WriteBookmarks\relax
\def\floatpagepagefraction{1}
\def\textpagefraction{.001}
\shorttitle{$J/\psi$-$J/\psi$ resonance}
\shortauthors{J.-M. Richard}

\title [mode = title]{About the $J\!/\!\psi$\,$J\!/\!\psi$ peak of LHCb: fully-charmed tetraquark?}                      

\author{Jean-Marc Richard}[
                         bioid=1,
                         orcid=0000-0001-6459-765X]
\cormark[1]
\fnmark[1]
\ead{j-m.richard@ipnl.in2p3.fr}

\address[1]{Institut de Physique des 2 Infinis, IN2P3 \& Universit\'e Claude Bernard,  4 rue Enrico Fermi, 69622 Villeurbanne, France}

%
%
%

%

\cortext[cor1]{Corresponding author}


\begin{abstract}
We discuss the interpretation of the $J\!/\!\psi$\,$J\!/\!\psi$ peak recently seen by the LHCb collaboration at CERN, and take the opportunity to review the experimental and theoretical state of the art for exotic hadrons. It is stressed that in constituent models, the states in the continuum require a dedicated treatment to single out resonances atop the continuum. 
\end{abstract}


\begin{highlights}
\item Exotic hadrons
\item LHCb experiment
\item Constituent quark model
\end{highlights}

\begin{keywords}
exotic hadrons \sep LHCb \sep quark model 
\end{keywords}

\maketitle

Since several decades, a particular attention has been devoted to hadrons with anomalous properties \cite{Montanet:1980te,Jaffe:2004ph,Richard:2016eis,Ali:2019roi,Brambilla:2019esw}. On the experimental side, the search has been very difficult and disappointing. For instance, in the early 60s, there has been some claims for baryons resonances with strangeness $S=+1$, the so-called $Z$ baryons, while the established baryons have either $S=0$ or $S<0$. But the $Z$ spectrum  was based on somewhat hazardous analyses of $K^+$-nucleon scattering data without spin observables, and it faded away. It reappeared under the name of $\theta$ pentaquark, but it was never firmly established. 

Another example is the  set of ``baryonium'' mesons~\cite{Montanet:1980te},  preferentially coupled  to baryon-antibaryon channels, and seen as peaks in antinucleon-induced reactions. A low-energy antiproton facility  was built at CERN in the 80s to study the baryonium, as a side product of the search for $W$ and $Z$ bosons in  high-energy antiproton-proton collisions, but none of the baryonium candidates was confirmed.  

From the above examples, and others, it became clear that the search for exotic hadrons requires intense beams, powerful detectors and sophisticated programs for the analysis. With the advent of charm and  beauty factories, i.e., high-intensity electron-positron colliders and  the newest hadronic colliders, more reliable data have been acquired, and, indeed, remarkable hadronic states have been discovered. 

The first and most emblematic state of this ``new generation'' is the $X(3872)$, a meson with hidden charm, but not fitting any radial and/or orbital excitation of charmonium, i.e., having a structure more complicated than a mere charm quark-antiquark pair~$c\bar c$. The $X(3872)$, first discovered with an electron-positron machine, has been confirmed in a variety of experiments. For a review, see, Ref.~\cite{Brambilla:2019esw}, with a critical discussion of other hidden-charm or hidden-beauty mesons. New pentaquark states have also been discovered by LHCb, again with hidden-charm content. The name ``pentaquark'' refers to baryons with a minimal content of four quarks and one antiquark. The LHCb pentaquarks have been described in Ref.~\cite{ZHANG20191119}.

Recently, the same LHCb collaboration has found a peak in the $J\!/\!\psi$\,$J\!/\!\psi$ mass spectrum \cite{Aaij:2020fnh}, with a mass of about 6900\,MeV, i.e., about 700\,MeV above the $J\!/\!\psi$-$J\!/\!\psi$ threshold, and a width of about 80\,MeV,  and this revitalized the studies about multiquark resonances made of heavy quarks and antiquarks. This is  a very important step in the exploration of the heavy hadrons, after the charmonium $c\bar c$ in 1974, the charmed hadrons $c\bar q$ and $cqq$ in the following years (here $q$ denotes a light quark),  the bottomonium $b\bar b$ in 1977, and then the mesons and baryons carrying beauty, the $B_c=b\bar c$ in 1996 at Fermilab, and the double-charm baryons $ccq$ in 2017 by LHCb. For a review of the salient contributions to the early part of this history, see, e.g., Ref.~\cite{Ezhela:317696}. 

So far, $J\!/\!\psi\,J\!/\!\psi$ is just the discovery channel. Obviously, the measurement of other properties is desirable. In particular, the decay into other pairs of charmonia, $\eta_c\eta_c$, $\eta_c\chi_J$, \dots would be crucial for identifying the quantum numbers and screening the  internal dynamics. The literature will very probably privilege either a four-quark state, or a ``molecular'' system of two charmonia with, perhaps, higher Fock components. There is an abondant literature on molecular hadrons, as reviewed, e.g., in Ref.~\cite{Brambilla:2019esw}. Its extension for a system of two charmonia is in order, with an interaction which is probably very short-ranged. Heavier molecules with double hidden-charm can be envisaged, made of excited charmonium states or a baryon-antibayron pair with double-charm, $\Xi_{cc}\bar\Xi_{cc}$~\cite{Meng:2017fwb}. 

On the theory side, a peak in such a hadron-hadron mass distribution stimulates new debates on an old problem: are there hadrons beyond the quark-antiquark mesons and the baryons made of three quarks? The molecular picture is supported by effective field theories: an effective Lagrangian is designed, inspired by QCD, and adapted to a specific sector of the hadron spectrum. At the quark-gluon level, the most advanced method is lattice QCD, in which the space-time is discretized~\cite{Brambilla:2019esw}.We shall restrict here to simple constituent models that already give a good insight on the problem of the existence of multiquarks. 
 
%
%
%
%
%

The simplest multiquark configuration consists of two quarks and two antiquarks. 
The light sector is somewhat the realm of chromomagnetism, and, in particular,
 the $qq\bar q\bar q$ states have been categorized according to the eigenvalues of the color-spin operator $\sum_{i<j}\llss{i}{j}$, where $\vec \sigma_i$ is the usual spin operator acting of the $i^\text{th}$ constituent, and $\tilde \lambda_i$ its analog for color. It has been realized that $S$-wave $qq\bar q\bar q$ configurations can compete with $P$-wave $q\bar q$ states in building the light scalar mesons~\cite{Jaffe:2004ph}. Moreover, the chromomagnetic interaction exhibits striking coherences that might lead to stable multiquarks in the dibaryon sector such as $uuddss$, or the sector of anticharmed baryons such as $\bar c uuds$, as reviewed, e.g., in Ref.~\cite{Richard:2016eis}.
 
 As the mass of the constituent quarks increases, the chromomagnetic term vanishes, as it contains for each pair a $1/(m_i\,m_j)$ factor, like in the Breit-Fermi interaction in atomic physics. Hence the spectroscopy becomes dominated by the chromoelectric interaction, and in the simplest model, the Hamiltonian reads
 \begin{equation}\label{eq:H}
  H=\sum_i \frac{\vec p_i^2}{2\,m_i}-\frac{3}{16}
  \sum_{i<j}\llcol{i}{j} \,v(r_{ij})~,
 \end{equation}
where the normalization is such that $v(r)$ is the quarkonium potential. This Hamiltonian, however simple as compared to the true QCD dynamics, should be treated with the care required for any few-body systems at the edge between stability and dissociation~\cite{Richard:2018yrm}. The bound-state part of $H$, if any,  is rather straightforward, as one can use standard variational methods involving a basis of normalizable trial wave-functions. It turns out that if $H$ is treated carefully, there is no bound state for equal masses, i.e., no $QQ\bar Q\bar Q$ tetraquark below the $Q\bar Q+Q\bar Q$ dissociation threshold. Above this threshold, specific methods are required to disentangle the genuine resonances from the artifacts due to using a finite basis of normalizable wave-functions. One of such methods, real scaling, has been applied by Hiyama et al.\ for pentaquarks. See, e.g., Ref.~\cite{Meng:2019fan} and refs.\  therein. Preliminary investigations indicate that within the Hamiltonian of Eq.~\eqref{eq:H}, resonances do exist for $cc\bar c\bar c$ in the mass range of the LHCb resonance. Experimentally, resonances decay by mere dissociation, while bound states use internal $Q\bar Q$ annihilation. 

Similar resonances are expected for the other  heavy configurations such as $bb\bar b\bar b$ or $bb\bar c\bar c$, which are, of course, difficult to observe with our present means. A special mention is deserved for $bc\bar b\bar c$, whose experimental signature is $\mu^+\mu^-\mu^+\mu^-$ corresponding to real of virtual $\Upsilon+J/\psi$. As $bc\bar b\bar c$ does not suffer from any Fermi statistics, the color part of its wave function can be rearranged when the quarks move, to capture the most favorable fluxes of gluon linking the quarks. Then the Hamiltonian of Eq.~\eqref{eq:H} can be modified to include a non-pairwise interaction for the confining part, which turns out more attractive; see, e.g., \cite{Richard:2016eis,Richard:2018yrm}.

A rather popular model, nowadays, describes the baryons and the multiquarks in terms of diquarks. This is a rather old idea, though the contributions of the pioneers \cite{Anselmino:1992vg} are not always acknowledged. If the diquark $\mathsf{ D}$ is taken as an effective degree of freedom, the difficulty is to relate the $\mathsf{D}\mathsf{\bar D}$ spectrum to  the quarkonium masses that enter the threshold. Now, the diquark sometimes appears as a short-cut to avoid the complications of the four-body problem, but it turns out that this approximation tends to overbind, with the risk of transforming resonances into bound states \cite{Richard:2018yrm}.

An interesting interplay of light-quark and heavy-quark 
physics is provided by the $QQ\bar q \bar q$ configuration with double heavy flavor, first discussed  decades ago (see  Ref.~\cite{Richard:2016eis}), and now regularly revisited. As compared to its threshold $Q\bar q+Q\bar q$, it benefits from two advantages: the strong $QQ$ attraction, a chromoelectric effect, and if $qq=ud$, a favorable chromomagnetic attraction. This will perhaps be the first stable multiquark, besides the deuteron. Amazingly, the chromo-electric effect is similar to the observation that the hydrogen molecule $ppe^-e^-$ is more stable than the positronium molecule $e^+e^+e^-e^-$, due to a favorable symmetry breaking. 

\subsection*{Conflict of interest}
 The author declares that he has no conflict of interest. 
 
\subsection*{Acknowlegments}
 It is a pleasure to thanks the experimental teams having tirelessly hunted for exotic mesons. I would also like to thank the community of exotic-hadron phenomenologists for many stimulating discussions along the years, in particular Alfredo Valcarce and Javier Vijande. I also thank M.~Asghar for useful comments. 
%

%

%
\bio{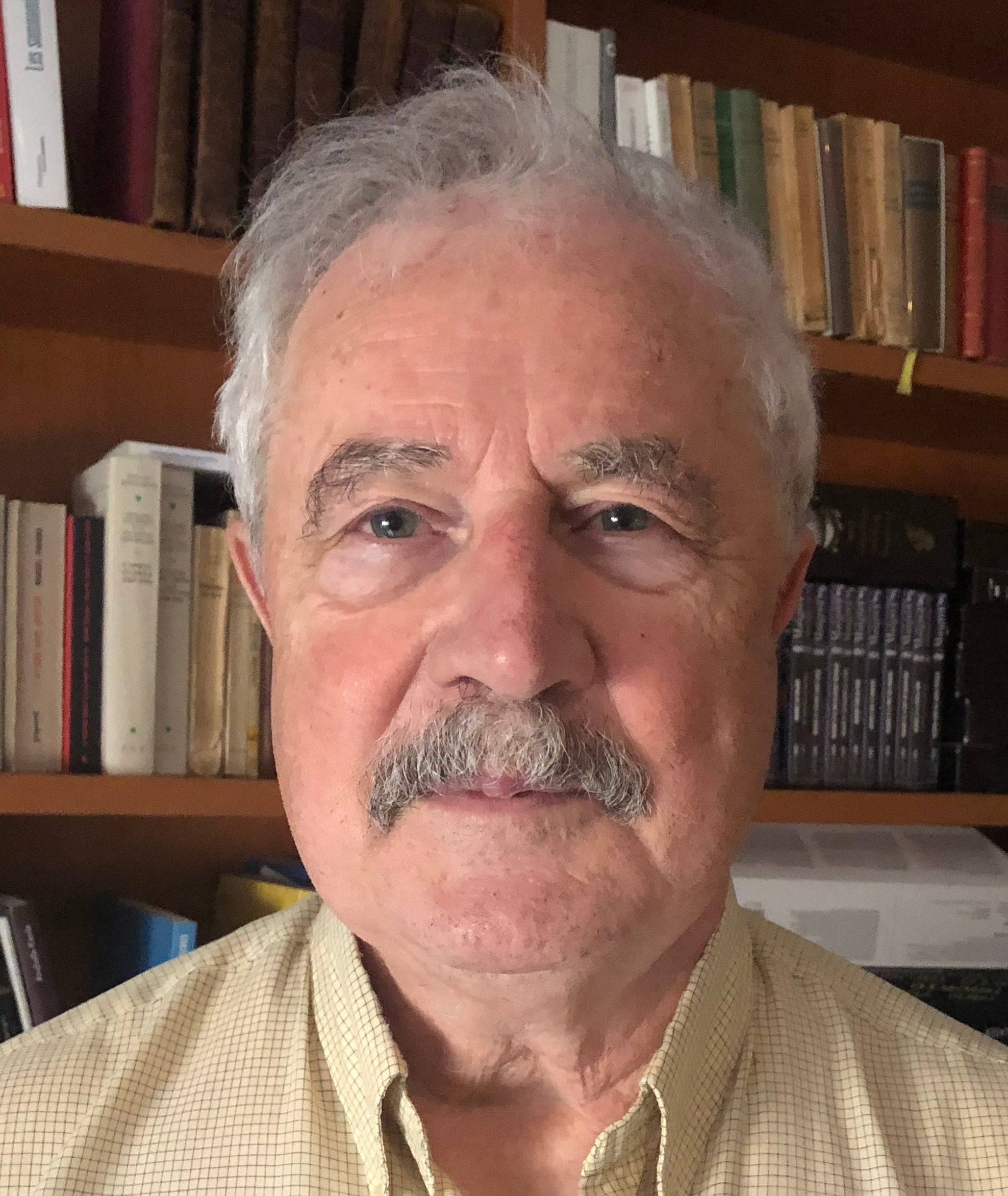}
Jean-Marc Richard received his doctorate in physics at the University of Paris in 1975. He started his carreer as Assistant-Professor in Orsay, and he is now Emeritus Professor at the University of Lyon. During sabbatical leaves, he visited Stony-Brook, Brookhaven, Aarhus, Heidelberg, Bonn, Trento, ILL  and CERN. His research focuses on nuclear physics, hadron physics, and few-body systems in atomic physics. 
\endbio
\end{document}